\documentclass[conference]{IEEEtran}
\IEEEoverridecommandlockouts
\usepackage{fancyhdr}
\usepackage[numbers]{natbib}
\usepackage{hyperref}
\usepackage{amsmath,amssymb,amsfonts}
\usepackage{algorithmic}
\usepackage{graphicx}
\usepackage{textcomp}
\usepackage{xcolor}
\usepackage{physics}

\def\BibTeX{{\rm B\kern-.05em{\sc i\kern-.025em b}\kern-.08em
    T\kern-.1667em\lower.7ex\hbox{E}\kern-.125emX}}
\begin{document}

\onecolumn
\null
\vfill
\begin{center}
    \huge{This article has been accepted for publication in the IEEE International Conference on Quantum Communications, Networking, and Computing 2025. This is the accepted manuscript made available via arXiv. }
\end{center}
\vfill
\normalsize{© 2025 IEEE. Personal use of this material is permitted. Permission from IEEE must be obtained for all other uses, in any current or future media, including reprinting/republishing this material for advertising or promotional purposes, creating new collective works, for resale or redistribution to servers or lists, or reuse of any copyrighted component of this work in other works.}

\twocolumn
\clearpage

\title{Combined Quantum and Post-Quantum Security for Earth-Satellite Channels}
 
\author{Anju~Rani$^{*}$, Xiaoyu~Ai$^{*}$, Aman Gupta$^{*}$, Ravi Singh Adhikari$^{*}$ and~Robert~Malaney$^*$ 
\thanks{$^*$Anju Rani, Xiaoyu Ai, Aman Gupta, Ravi Singh Adhikari, and Robert Malaney are with the School of Electrical Engineering and Telecommunications, University of New South Wales, Sydney, Australia.}
}

\maketitle
\begin{abstract}
Experimental deployment of quantum communication over Earth-satellite channels opens the way to a secure global quantum Internet. In this work, we present results from a real-time prototype quantum key distribution (QKD) system, which entails the development of optical systems including the encoding of entangled photon pairs, the development of transmitters for quantum signaling through an emulated Earth-satellite channel, and the development of quantum-decoding receivers. A unique aspect of our system is the integration of QKD with existing cryptographic methods to ensure quantum-resistant security, even at low-key rates. In addition, we report the use of specially designed error-reconciliation codes that optimize the security versus key-rate trade-off. Our work demonstrates, for the first time, a deployment of the BBM92 protocol that offers both post-quantum security via the advanced encryption standard (AES) and quantum security via an entanglement-based QKD protocol. If either the AES or the QKD is compromised through some adversary attack, our system still delivers state-of-the-art communications secure against future quantum computers.
\end{abstract}


\section{Introduction}
The security of  QKD is fundamentally rooted in the principles of quantum mechanics. In theory, QKD ensures secure communication between two legitimate users, even in the presence of a malicious eavesdropper. Unlike classical cryptography, which relies on the complexity of mathematical algorithms for security, QKD exploits quantum attributes such as the no-cloning theorem, the indistinguishability of non-orthogonal quantum states, or quantum entanglement to provide information-theoretic security (\textit{i.e.,} provable security guarantees that are independent of the resources held by the adversary). QKD has achieved significant milestones in global intercontinental deployment, including high secret key generation rates~\cite{islam2017provably,yuan201810}, long-distance quantum communication~\cite{pittaluga2021600,xie2022breaking}, photonic integration~\cite{sibson2017chip,bunandar2018metropolitan}, and the development of terrestrial and satellite-based quantum networks~\cite{chen2021integrated}.

Since the introduction of the first QKD protocol (BB84~\cite{bennett2014quantum}), various discrete-variable (DVQKD)~\cite{liu2023experimental} and continuous-variable (CVQKD)~\cite{grosshans2002continuous} protocols have been developed. However, QKD implementations encounter inherent challenges such as non-ideal light sources, imperfect detection, and lossy channels. A major limitation within QKD is the rate-loss trade-off, which constrains both the distance and the key rate due to channel loss. This limitation can be addressed using quantum relays~\cite{collins2005quantum} or trusted relays~\cite{stacey2015security}. However, quantum relays are based on quantum memory and high-fidelity entanglement distillation, technologies that are beyond current capabilities. Trusted relays, while conceptually simpler, assume a secure intermediary between users, an assumption that is difficult to verify in practice. Although CVQKD can improve the key rate in some circumstances, its transmission distance over free space remains significantly shorter than that of DVQKD (at least in reported deployments). Taken together, these issues suggest that point-to-point DVQKD is the best deployment for Earth-satellite channels, at least in the near term.

The future of cryptographic architecture is most likely a modular one that allows easy replacement (or upgrade) of any cryptographic primitive if a vulnerability is discovered in it. QKD is a key-sharing primitive that can be incorporated into such an architecture rather than replacing existing primitives. An architecture with QKD alone requires a one-to-one key-to-message bit ratio, which is often impractical for large-scale data encryption. Currently, the required key rate for bandwidth-intensive communication is high compared to what QKD can provide in real-time - continuous generation of keys at the required rates is technologically challenging. Alternatively, post-quantum cryptography (PQC) primitives serve as a software-based option to more expensive QKD implementations~\cite{bernstein2017post}. However, PQC does not provide information-theoretic security but remains, as far as we currently know, computationally secure against a quantum adversary (\textit{i.e.}, quantum-resistant security). Therefore, a modular cryptographic architecture that incorporates both QKD and PQC cryptographic primitives  could be beneficial. In case of a cryptographic failure or newly discovered vulnerability in either of the primitives, such an architecture could continue to provide, as a minimum, quantum-resistant security.

The main contribution of this work is the first QKD-PQC implementation that sequentially combines a QKD primitive based on BBM92 with a primitive based on AES with a 256-bit key length (AES-256). We note AES-256  provides quantum-resistant security. As we discuss later, this sequential combination of the primitives can offer an advantage relative to other combined QKD-PQC architectures, especially in the Earth-satellite channels - the communication scenario of main interest to us. 

The remainder of this paper is structured as follows. Section~\ref{sec:Theory} provides the theoretical background of the BBM92 protocol and introduces the PQC used for an additional layer of security. Section~\ref{sec:Experiment}  details the experimental implementation, and Section~\ref{sec:RnD} presents the results. Finally, Section~\ref{sec:Conclusion} concludes with some remarks on our findings. 

\section{\label{sec:Theory}Background}
\subsection{Quantum Key Distribution}
In our system, we produce QKD keys in real time using an entangled version of QKD, namely BBM92, which is information-theoretic secure  and applicable for long distances~\cite{liorni2019satellite}. BBM92 is equivalent to the prepare and measure system of BB84 when the sender measures their photons (their half of the pairs of entangled photons) in mutually unbiased bases.  More generally, in BBM92 the source of the photons is an intermediary device sending one photon from each entangled pair to two separate receivers. The channel between these two receivers then becomes a quantum channel. Based on measurements at the receivers and some subsequent information reconciliation followed by privacy amplification, a random number between them can be shared in an information-theoretic secure manner.

However, the use of an entangled source can provide other benefits to the network in which it is located, such as entanglement distribution, enhanced synchronization, and a pathway to device-independent QKD, the holy grail of secure communications. As we shall see in the following implementation, a major source of bit error in the BBM92 protocol is the time-tagging errors of the photons arriving at the detector. However, this is an issue in most QKD implementations. Once a key is derived within QKD it can then be used as a one-time-pad (OTP) for encryption via an XoR operation on the message to be sent. This encrypted message delivery is information-theoretic secure, provided that the size of the message is less than or equal to the size of the available QKD key. One important point in QKD is that it is always assumed that a pre-shared secure key is already in place between the sender and receiver. In this sense QKD is a \textit{key-growing} algorithm, it does not create an encryption key from nothing. As we describe next, some part of this pre-shared key will also be used within our PQC solution.

\begin{figure}[h]
		\centering		
        \includegraphics[width=01\linewidth]{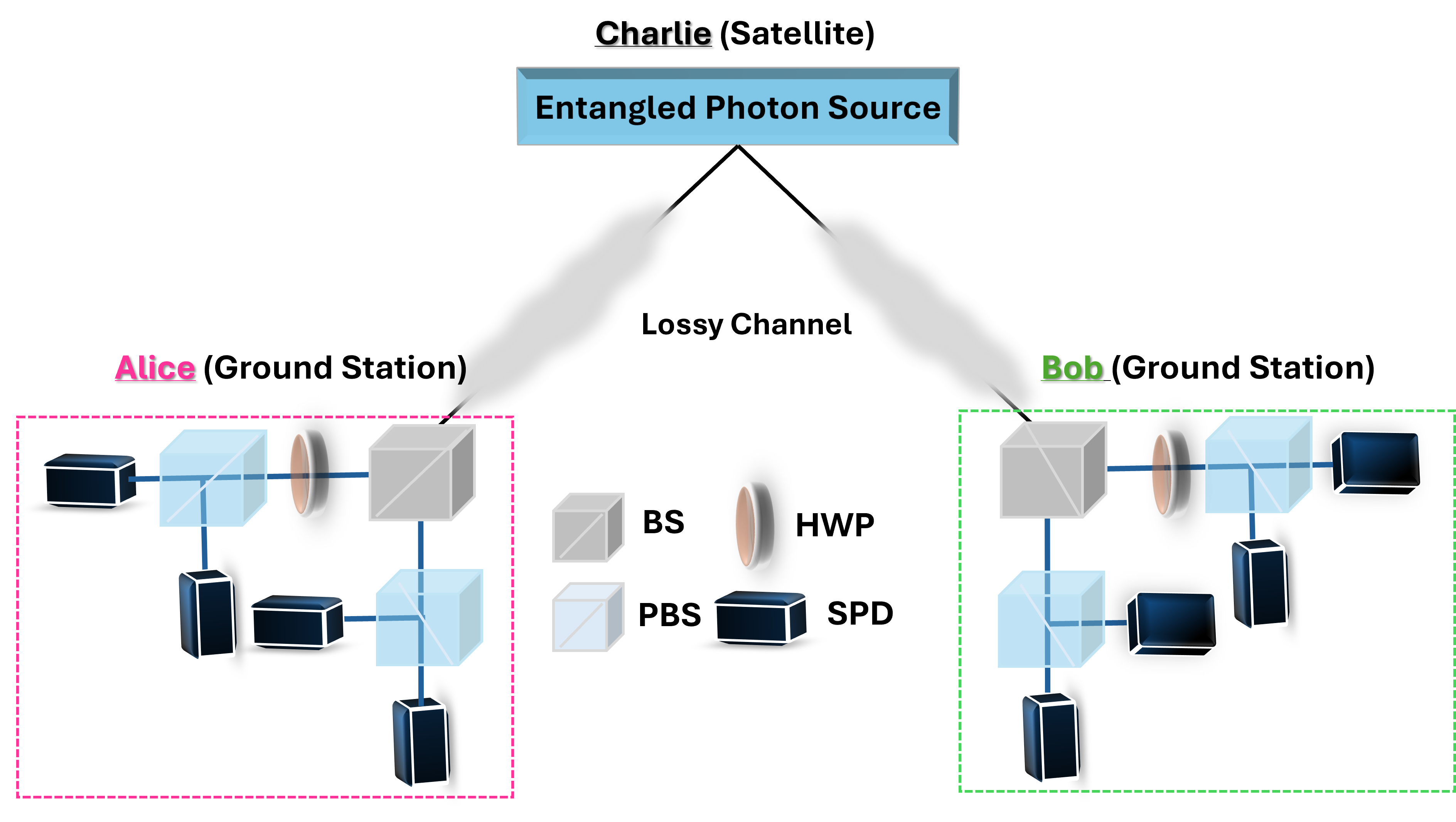}
		\caption{Our entanglement-based satellite quantum communication system. HWP: Half-wave plate; PBS: Polarizing beam splitter; SPD: Single-photon detector; BS: Beam splitter.}
 \label{fig:1}
\end{figure}

A schematic of BBM92 as implemented on a satellite-to-Earth channel is given in Fig.~\ref{fig:1}. Here, the ground stations receiving the photons are referred to as the ubiquitous Alice and Bob, and the satellite as Charlie. This is the setup we attempt to emulate here in a laboratory implementation - more details on the experimental implementation are provided later.
\subsection{Post Quantum Cryptography}
In 2016, the National Institute of Standards and Technology (NIST) announced a competition to submit post-quantum secure key exchange mechanisms~(KEMs) and digital signature schemes, of which CRYSTALS-Kyber for KEM and CRYSTALS-Dilithium, FLACON, and SPHINCS for digital signatures were chosen winners. All of these algorithms are recommended replacements for current asymmetric key cryptographic primitives. However, the symmetric key AES, with a minimum 256-bit key length, remains the fastest post-quantum secure solution for encryption and decryption of large datasets, and this AES-256 we use in this work. 

Since all PQC algorithms are computationally secure rather than provably secure, there will always be an inherent risk that they will be proven insecure in the future. A PQC algorithm (a form of CRYSTALS-Kyber) announced in 2022 was broken in July 2023~\cite{dubrova2023breaking}. Therefore, there is an emerging focus on future cryptographic architectures in which multiple cryptographic primitives are employed. These aim to provide quantum-resistant security, modular enough to easily upgrade a primitive whenever a vulnerability or attack is discovered, and robust enough to continue ensuring post-quantum security at the very least. 

\subsection {Combined Systems} 
The future of cryptographic architecture is most likely the one in which the choice of key share, whether to use QKD or any post-quantum primitive (or some combination) would depend largely on the security requirement of the data and the rate of secret key generation (\textit{i.e.,} QKD key generation). Therefore, recent work focuses on a modular architecture that combines all available cryptographic primitives and future primitives to allow easy transition and handling of cryptographic failures and newly discovered vulnerabilities more effectively. Two primary directions of work that combine QKD with PQC primitives focus on integrating PQC into the QKD key reconciliation step to address vulnerabilities in the classical communication part of QKD or to perform channel authentication in the key reconciliation step in the absence of a pre-shared key~\cite{djordjevic2020joint,pastushenko2023improving}. A third direction focuses on the generation of hybrid keys by combining the keys from both PQC and QKD primitives to provide modularity in cryptography and to improve the individual security of the keys generated from each primitive~\cite{garms2024experimental,bruckner2023end}.

\section{\label{sec:Experiment}Experimental Setup}
In our system, following on from the QKD encryption, we perform an additional encryption using AES-256 at the sender (Alice), with a corresponding decryption at the receiver (Bob). We do these additional operations not only to the message that is to be sent but also to all syndrome messages that form part of the information reconciliation within QKD. Although not implemented here, \textit{all} classical messaging that is part of the QKD protocol can undergo these additional classical encryption/decryption operations. As stated earlier, AES with pre-shared keys of at least 256 bits is considered computationally secure against known quantum computer-based attacks~\cite{bonnetain2019quantum}). The 256-bit keys used for AES-256  are fixed keys chosen from a part of a key pre-shared \textit{a priori} between Alice and Bob (recall that a separate part of this pre-shared key is also required for QKD). If the QKD keys are less than the message size, then the message is sent directly for AES-256 encryption without waiting for the accumulation of QKD keys. In this way, our system ensures that the data are secure against a quantum adversary, even if a new vulnerability or attack is discovered in the implementation of either the QKD or the AES (but not a successful attack on both). Furthermore, since we do not generate hybrid keys, as proposed in similar works~\cite{garms2024experimental,dowling2020many}, our protocol is without the burden of equal-length QKD and PQC keys needed for the generation of hybrid keys, and our implementation avoids the compositional security issues that could arise for hybrid keys~\cite{ricci2024hybrid}. We discuss these differences in more detail later.

\begin{figure*}
		\centering		
        \includegraphics[width=0.7\linewidth]{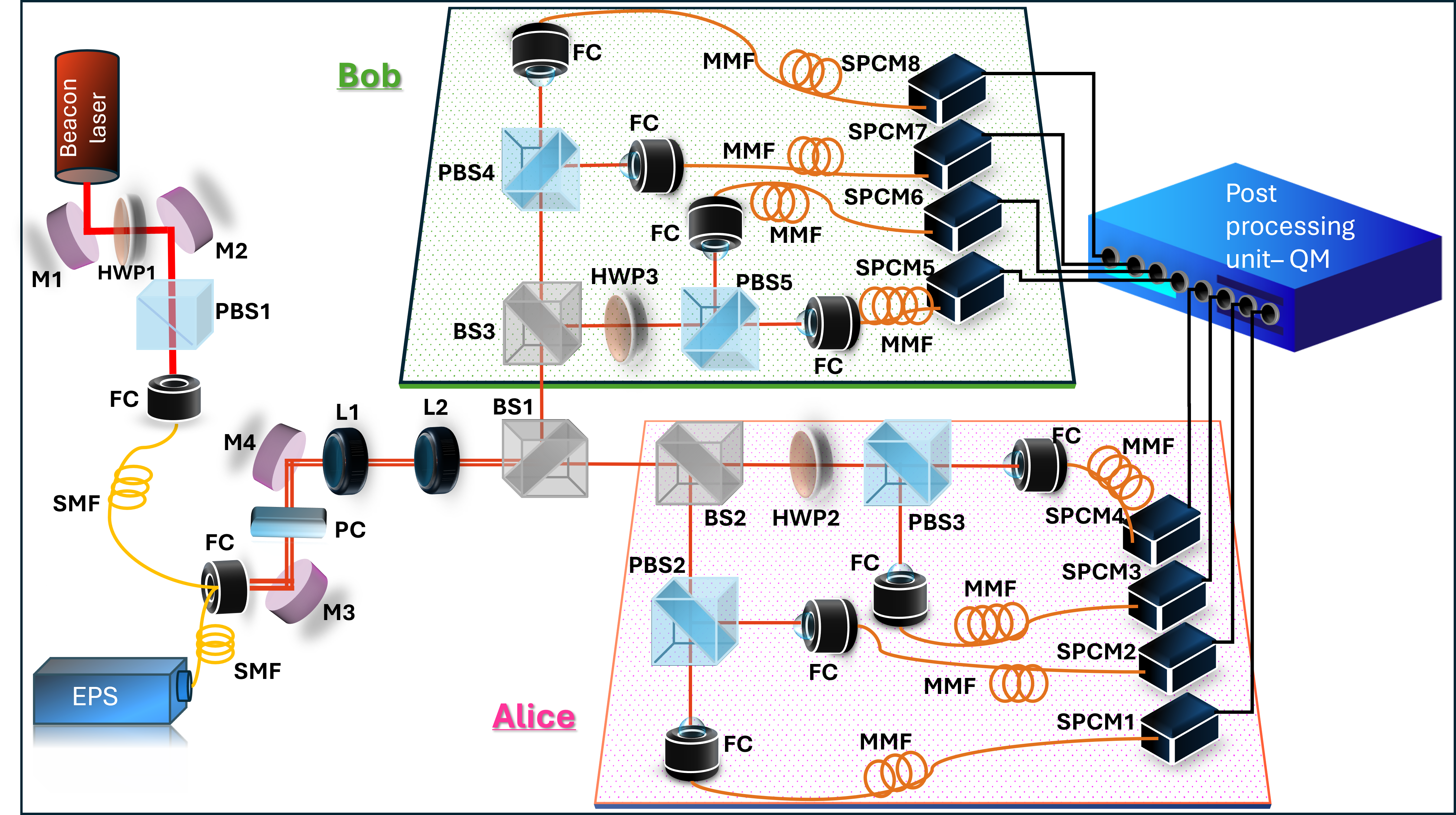}
		\caption{The experimental setup for the BBM92 protocol. EPS: Entangled photon source; M: Mirror; HWP: Half-wave plate; PBS: Polarizing beam splitter; FC: Fiber coupler; SMF: Single-mode fiber; BS: Beam splitter; PC: Polarization correction optics which is a combination of a quarter-wave plate - half-wave plate - quarter-wave plate; L: Lens; MMF: Multi-mode fiber; SPCM: Single-photon counting module; QM: Quantum machine (post-processing unit).}
 \label{fig:2}
\end{figure*}

Our alignment and optimization for the experiment involve the following steps: A beacon laser (HeNe) operating at 633~nm is used alongside the entangled photon source (EPS) for initial alignment, as shown in Fig.~\ref{fig:2}. The beacon laser is coupled to a fiber coupler (FC) via a single-mode fiber (SMF), and its beam is collimated at the output using a lens combination (L1 and L2). The beacon laser is then utilized to align the channel path and the detection setup for Alice and Bob. 

The EPS (Qubitekk) is fiber-coupled and generates entangled photon pairs — the signal and idler — at $810$~nm. Its design is based on a Hong-Ou-Mandel interferometer, utilizing a type II nonlinear ppKTP crystal for photon pair generation. The signal and idler photons are orthogonally polarized. The entangled state produced by the source is
\begin{equation}
    \ket{\psi} = \frac{1}{\sqrt{2}}(\ket{H}_{s}\ket{V}_{i}+\ket{V}_{s}\ket{H}_{i}),
\end{equation}
where $\ket{H}_{s}$ and $\ket{V}_{s}$ represent horizontally and vertically polarized photons, respectively, in the signal  (likewise idler with subscript $i$). The source can be tuned to achieve different visibilities depending on the input optical power levels. After coarse alignment using the beacon laser, the fiber is switched to the EPS. The signal and idler photons co-propagate through the fiber before being launched into free space via a fiber coupler. Polarization drift occurs due to fiber coupling, affecting the polarizations of the signal and idler photons. To compensate for this polarization drift, we use a quarter-wave plate – half-wave plate – quarter-wave plate 
in combination adjusting the angle of the wave plates to correct for the drift of polarization. A 50:50 beam splitter then separates the photon pairs, distributing them between Alice and Bob (see Fig.~\ref{fig:2}).

The photons received at Alice and Bob are measured by projecting their polarization states onto four basis states using a combination of half-wave plates (HWPs) and polarizing beam splitters (PBSs), followed by single-photon detectors (SPDs). We used an Excelitas single-photon counting module (SPCM) with a dark count rate of 100~per second. Alice and Bob have identical detection setups, each featuring a 50:50 beam splitter that randomly selects the basis for projection measurements, as shown in Fig.~\ref{fig:2}. A $\{H, V\}$ basis measurement is performed using PBS2 (Alice) and PBS4 (Bob), while a $\{D, A\}$ basis measurement is performed using HWP2 and PBS3 (Alice) and HWP3 and PBS5 (Bob). The photons are then coupled into multi-mode fibers and detected by SPCMs. The detected photons and their arrival times are recorded for post-processing and the subsequent generation of the secret (QKD) keys, as detailed next.

\subsection{Post-processing}
\label{sec:post-processing}
Post-processing plays a crucial role in generating QKD keys from detection events at the SPCMs. In our setup, the post-processing hardware consists of a time-tagging device (manufactured by Quantum Machine) shared between Alice and Bob. Our setup utilizes a total of eight independent input channels. Four of these channels are connected to Alice’s four SPCMs, while the remaining four are connected to Bob’s SPCMs. The time-tagging device monitors all eight channels. In this experiment, we assume that Alice and Bob's time-tagging devices are perfectly synchronized in their detection periods. However, in real-world deployments, Alice and Bob would each require their own time-tagging devices due to their physical separation. Ensuring accurate synchronization between these devices in such scenarios presents a technical challenge, but one that can be achieved to an accuracy that provides non-zero key rates~\cite{chen2021integrated}.
\par
\begin{figure*}
		\centering		
        \includegraphics[width=0.8\linewidth]{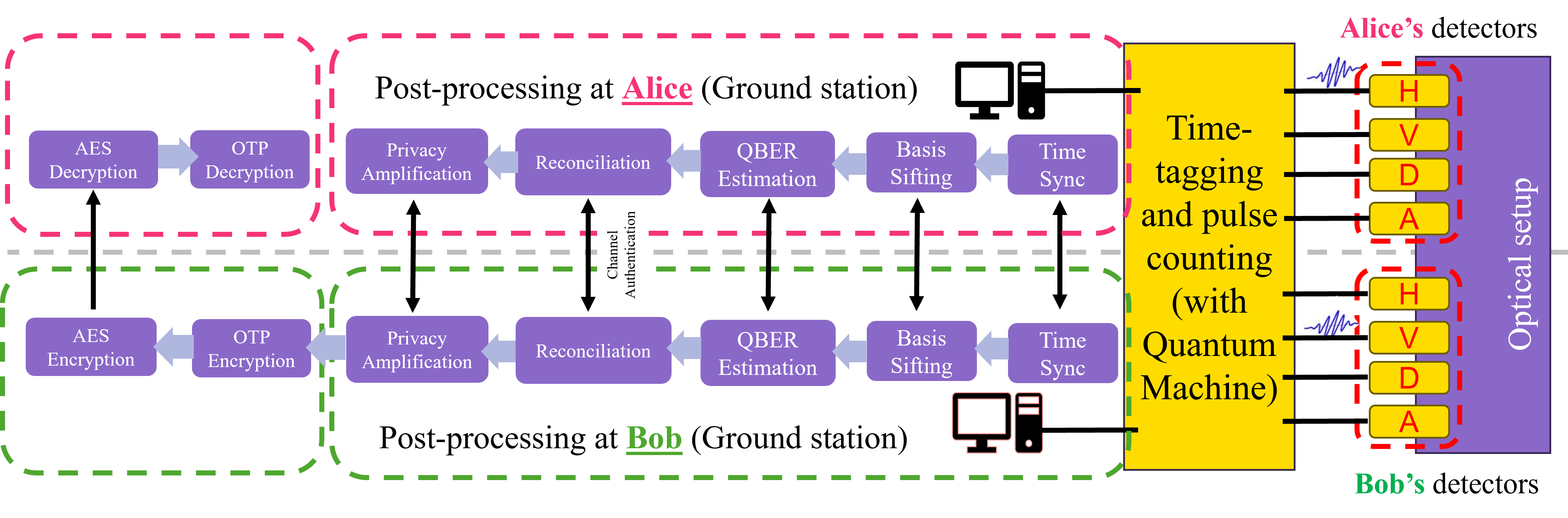}
		\caption{A high-level diagram for the QKD-PQC post-processing hardware setup.}
 \label{fig:post-processing}
\end{figure*}

 Alice and Bob independently use a personal computer to perform the post-processing steps, including interactive time-domain filtering, basis sifting, QBER estimation, reconciliation, and privacy amplification, as illustrated in Fig.~\ref{fig:post-processing}. We detail the steps of the post-processing.

\paragraph{Photon Time-tagging} \label{time-tagging} Once the system is turned on, the time-tagging device monitors all eight input channels within each detection period, which is triggered by a shared clock and set to $50$~ns in our experiment. During each detection period, if a digital pulse reaches any of the eight input channels, the time-tagging device assigns a positive integer as the time tag (representing the ceiling of the pulse arrival time in nanoseconds relative to the start of the detection period). If no pulses are detected, the time-tagging device will assign $-1$ as the time tag. The assigned time tags for the current detection period are then stored in separate arrays: one for Alice's SPCM (defined as $\mathbf{T_a}$) and another for Bob's SPCM (defined as $\mathbf{T_b}$). These arrays are then processed on their respective computers in the next step.

\paragraph{Time-domain Filtering} At this stage, Alice and Bob exchange their respective time-tag arrays, $\mathbf{T_a}$ and $\mathbf{T_b}$, via classical communications, and apply time-domain filtering. Specifically, they compare each corresponding time tag $t^i_a \in \mathbf{T_a}$ and $t^i_b \in \mathbf{T_b}$ and retain a specific tag in $\mathbf{T_a}$ (and the corresponding tag in $\mathbf{T_b}$) for the next post-processing step only if the following two conditions are met:
\begin{itemize}
    \item Condition 1: exactly one of the elements in $\mathbf{T_a}$ is a positive integer and exactly one of the elements in $\mathbf{T_b}$ is a positive integer. 
    \item Condition 2: the absolute difference between the positive time tag in $\mathbf{T_a}$ and $\mathbf{T_b}$ is less than or equal to a preset coincidence window, $t_c$. We set $t_c = 1$ ns in our experiment.
\end{itemize}

\par The first two post-processing steps ($a$ and $b$) are important in free-space optical deployments. Ambient light can cause unintended photon detections at random times, leading to an increased QBER. However, by precisely recording detection times and selecting only events where the time difference between Alice and Bob detections falls within $t_c$, such unwanted detections can effectively be filtered out. 

\paragraph{Bit Mapping} In this step, for each remaining time tag in $\mathbf{T_a}$ (and the corresponding time tag in $\mathbf{T_b}$), Alice determines two bits, that map to which one of her four SPCMs registered the pulse. The convention used by Alice for this bit mapping is $H\rightarrow0$, $V\rightarrow1$, $D\rightarrow1$, and $A\rightarrow0$.
The first mapped bit (corresponding to $H$ or $V$) is then stored in a bit array, $\mathbf{k_a}$, and the second bit (corresponding to the basis used) is stored in an array, $\mathbf{b_a}$. Bob performs a similar procedure, but follows the sequence of bit mappings: $H\rightarrow1$, $V\rightarrow0$, $D\rightarrow0$, and $A\rightarrow1$.
His first mapped  bit is stored in $\mathbf{k_b}$ and the second bit in $\mathbf{b_b}$. Steps $a - c$ are repeated until the length of $\mathbf{k_a}$ reaches a target length, $N_{\text{raw}}$, after which Alice and Bob proceed to the next step (for most of our experiments, we set $N_{\text{raw}} = 40,000$).

\paragraph{Base Sifting} In this step, Alice and Bob exchange their $\mathbf{b_a}$ and $\mathbf{b_b}$ via classical communications. Alice then compares $\mathbf{b_a}$ and $\mathbf{b_b}$ and discards her bits in $\mathbf{k_a}$  where the corresponding basis labels do not match.  Bob follows the same procedure for $\mathbf{k_b}$ and ensures that only bits measured in the same basis are retained. The length of the remaining $\mathbf{k_a}$ and $\mathbf{k_b}$ will be $\frac{N_{\text{raw}}}{2}$.

\paragraph{QBER Estimation} In this step, Alice randomly selects a subset (with the size of $\frac{N_{\text{raw}}}{4}$) of bits from $\mathbf{k_a}$ along with their corresponding indices, and sends them to Bob. Bob then compares these received bits with the bits at the same indices in $\mathbf{k_b}$ and estimates the QBER (denoted by $Q$). The QBER is calculated as the ratio of the number of bit discrepancies found in the comparison to the total number of bits used for estimation. Next, Bob discards the bits used in the comparison and informs Alice of the estimated QBER. Alice then also removes the selected bits from $\mathbf{k_a}$. We denote the length of $\mathbf{k_a}$ and $\mathbf{k_b}$ by the end of this step as $N$. 
\paragraph{Error Correction} In this step, Alice and Bob will correct the discrepancies between $\mathbf{k_a}$ and $\mathbf{k_b}$ via a one-way classical and authenticated channel with a pre-shared Low Density Parity Check (LDPC) matrix, $\mathbf{H}_{M\times N}$, where $M = \lceil(1-R_c) N\rceil$ and $R_c$ is the rate of the LDPC code. The specific matrix\footnote{The construction of $\mathbf{H}_{M\times N}$ follows a two-step process. First, for a given $R_c$, we determine the degree distributions of the optimized LDPC matrix for the correction of the error of the discrete variable QKD by referring to Table.~1 of~\cite{elkouss2009efficient}. Second, for a given $N$ and the determined degree distributions, we construct $\mathbf{H}_{M\times N}$ with the Progressive Edge Growth algorithm~\cite{xiao2004improved}.} we use here is designed to optimize the security versus key-rate trade-off and as far as we are aware this work represents the first use of these specific codes in an actual implementation of an entanglement based QKD system.
 Then Bob calculates his syndrome array, $\mathbf{s_b}$, by performing $\mathbf{s_b} = \mathbf{H}_{M\times N} \mathbf{k_b}$. For authentication, Bob computes the MAC tag ($\mathcal{T}$) for the syndrome $\mathbf{s_b}$ using the Wegman-Carter-MAC\footnote{Wegman-Carter-MAC uses a one-time MAC signature as a polynomial hashing algorithm.  The advantage of using this MAC in QKD is that it consumes only $\log_2(M)$ bit keys from the previous QKD session, here $M$ is $length(Syndrome)$. To generate a one-time signature using two 128-bit fixed pre-shared, a hash is generated as: $Sign (k_1, \mathbf{s}) = (\mathcal{P}_{\mathbf{s}}(k) + a) \text{ mod }q$, where, $k_1 = (k,a)$ (128 bit fixed keys), $\mathbf{s} = (\mathbf{s}(1),\cdots,\mathbf{s}(M))$, $P_{\mathbf{s}}(k) = \mathbf{s}(M).k^M+\cdots+\mathbf{s}(1).k^1$, $q$ ($=2^M-1$ in this case) is a large prime. The tag is then generated as $WC((k_1,k_2),\mathbf{s}) = k_2 \text{ XoR } Sign (k_1, \mathbf{s})$, where $k_2$ is the OTP key (of size $\lceil \log_2(M) \rceil$ bits) from the previous QKD session.}. Bob uses the AES encryption algorithm, $\mathcal{E}_p$ ($\mathcal{D}_p$ is the corresponding decryption algorithm), with a part of the pre-shared key to encrypt the syndrome, leading to an output $\mathcal{E}_p(\mathbf{s_b})$. Bob sends \{$\mathcal{E}_p(\mathbf{s_b})||\mathcal{T}$\} to Alice via a classical communication channel (`$||$', represents the concatenation operation). After receiving \{$\mathcal{E}_p(\mathbf{s_b})||\mathcal{T}$\}, Alice decrypts the syndrome to obtain $\mathbf{s_b}$ and verifies the MAC tag and if the verification is successful she starts the error-correction with a belief-propagation decoder that takes $\mathbf{H}_{M\times N}$, $\mathbf{s_b}$ and $\mathbf{k_a}$ as the inputs, else if the verification fails then they abort the current session and restart from step $a$. The error correction will succeed if the belief-propagation decoder returns a a modified $\mathbf{{k}_a}$, referred to as $\mathbf{\hat{k}_a}$, that has the same syndrome with $\mathbf{s_b}$. 
\par Finally, if the LDPC decoder does not fail, Alice and Bob will move to the next step to generate QKD secret keys. Otherwise, they abort and restart the protocol from step $a$.
\paragraph{Privacy Amplification} In this step, Alice and Bob extract secret keys based on the estimated $Q$. First, Alice calculates the final QKD key rate in bits per pulse, $r$, using the asymptotic analysis of BBM92~\cite{cai2009finite}:
\begin{equation}
\label{eq:key_rate}
    r = 1-h(Q) - \text{leak}_{EC}\,,
\end{equation}
where $h(x) = -x\log (x) - (1-x)\log (1-x)$ is the binary entropy function and $\text{leak}_{EC}$ is the ratio of information disclosed to Eve during the error correction step and the length $N$. Then, Alice creates a Toeplitz matrix, $\mathbf{T}$, with a dimension of $\lceil rN\rceil \times N$ following the procedure described in~\cite{bourgoin2015experimental}. Next, Alice obtains her QKD secret key, $\mathbf{k^\prime_a}$ by performing $ \mathbf{k^\prime_a} = \mathbf{T}\mathbf{\hat{k}_a}$. Alice sends $\mathbf{T}$ to Bob so that he can obtain his secret QKD key $\mathbf{k^\prime_b}$ by performing $\mathbf{k^\prime_b} = \mathbf{T}\mathbf{k_b}$. 
\par The above key-rate analysis will hold exactly as the key length approaches infinity. Of course, such an assumption does not hold in any real deployment. Errors will be introduced for realistic key lengths due to estimates of the QBER possessing inaccuracies due to finite sampling. However, previous analyses have shown that for key lengths greater than 1 million bits and for QBER mean values in the range used here, the finite key effect will be negligible~\cite{unswuts}. Future work will fully account for finite sampling effects, but in the meantime we can take the rates listed here as upper limits.

\paragraph{\label{Enc-Dec}Encryption and Decryption} After the fresh secret keys have been generated, our model goes into encryption mode. Assuming that Alice wishes to send the data to Bob; if $length(\mathbf{Z})\leq length(\mathbf{k_a'})$, where $\mathbf{Z}$ are the data bits, then the encryption of OTP with secret keys results in $\mathcal{E}_q(\mathbf{Z}) = \mathbf{Z} \text{ XoR } \mathbf{k_a'}$ ($\mathcal{D}_q$ is the corresponding decryption algorithm). The encrypted message is then passed for a second encryption with AES using pre-shared keys, $\mathbf{k_p}$. The output $\mathbf{Z'} = \mathcal{E}_p(\mathcal{E}_q(\mathbf{Z}),\mathbf{k_p})$ is sent to Bob which is then decrypted using the corresponding reverse decryption algorithms and keys,
$\mathbf{Z} = \mathcal{D}_q(\mathcal{D}_p(\mathbf{Z'},\mathbf{k_p}),\mathbf{k_b'})$. If $length(\mathbf{Z})> length(\mathbf{k_a'})$, then we either wait for more QKD keys to be collected or we skip the OTP encryption to perform the AES-256 encryption and the corresponding decryption directly at Bob's end. Two systems with 16 GB RAM and a 2.90 GHz clock speed GPU were used for Alice and Bob. We perform the entire system run, from the collection of fresh keys to the encryption-decryption, and we analyze the time consumed for different message sizes, suggesting a negligible time for the encryption-decryption step compared to the time consumed by the accumulation of QKD keys. The same AES-256 processes were also applied to all syndrome messages that were sent between Alice and Bob as part of the reconciliation process. As discussed later, this has some advantages under certain attacks. In future implementations, we will also incorporate encryption of \textit{all} classical reconciliation messages using AES. This additional use of AES-256 will follow the description just outlined; no additional functionality will be needed.  

\subsection{\label{sec:emulation}Emulating Earth-Satellite Channels} 
We are interested in emulating our laboratory experiment for satellite communication applications. The laboratory test can be emulated as a satellite-based communication system, where the EPS is presumed to be onboard a low Earth Orbit (LEO) satellite. The EPS distributes the pairs of entangled photons to two optical ground stations, representing the communication parties, Alice (ground station~1) and Bob (ground station~2)~\cite{li2024microsatellitebasedrealtimequantumkey}. This downlink communication model offers higher efficiency compared to uplink communication due to lower signal attenuation in the downlink than in the uplink~\cite{1263786}. Indeed, in the downlink channel most propagation is through a vacuum before being disturbed by the atmosphere. The loss of signal photons over the atmosphere is complex and involves many factors including absorption, scattering, diffraction, pointing loss, and a host of other factors related to turbulence effects.

However, here we make some assumptions to simplify our modeling. We assume that the total channel link loss (in dB) is given by~\cite{joarder2025entanglementbasedquantumkeydistribution} $A_{link} =A_{atm} + A_{fs}$, where $A_{atm}$ accounts for losses due to atmospheric absorption and scattering, and $A_{fs}$ accounts only for beam diffraction, optical losses in the telescopes, and pointing loss. 
The value of $A_{fs}$ under our assumptions is given by~\cite{1263786} 
\begin{equation}
    A_{fs} =10\log_{10} \left(\frac{L^{2} \lambda^{2}}{{D_{T}}^{2} {D_{R}}^{2} T_{T} T_{R} (1-P_{l})}\right),
\end{equation}
where $L$ denotes the link distance, $\lambda$ is the operating wavelength, $D_{T}$ and $D_{R}$ denote the diameters of the transmitting and receiving telescopes, respectively, $T_{T}$ and $T_{R}$ represent the transmission efficiency of transmitting and receiving telescopes, respectively,  and $P_{l}$ is the pointing loss. 
\begin{figure}[h]
		\centering		
        \includegraphics[width=01\linewidth]{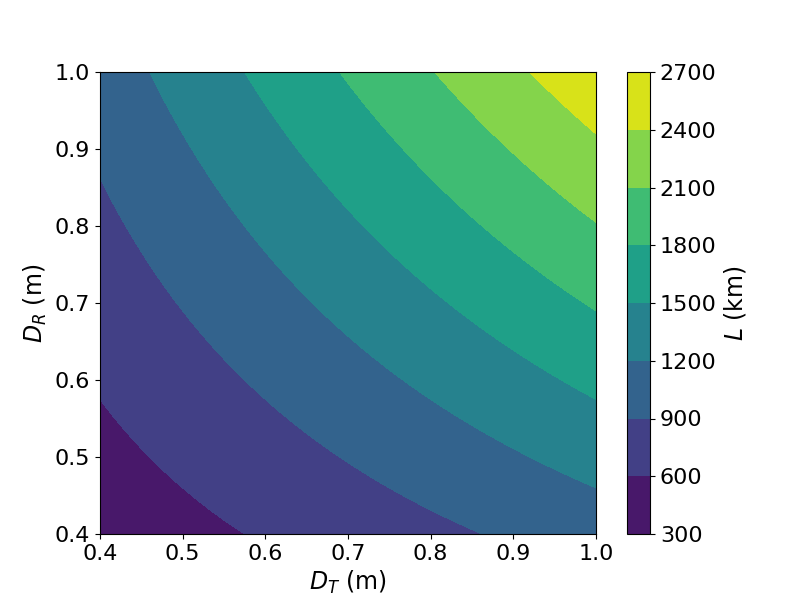}
		\caption{Plot of link distance $L$ as a function of transmitter ($D_{T}$) and receiver ($D_{R}$) telescope diameters for fixed link loss of $10.3$~dB, $\lambda=800$~nm, $P_{l}=0.2$, and $T_{T}=T_{R}=0.8$.}
        \label{fig:4}
\end{figure}
In our experimental optical setup, the link losses are set to  be ${A_{link}}= 10.3$~dB (channel losses of 4.3~dB and detection losses of 6.0~dB). This loss value is consistent with the anticipated satellite-to-Earth channels for reasonable telescope sizes. To show explicitly the relationship between telescope sizes and propagation distance we set the following parameters. 
We set $A_{atm}=1$~dB (which implies favorable atmospheric conditions), 
$\lambda=800$~nm, $T_T=T_R=0.8$, and $P_{l}=0.2$. Then, we determine for which $D_R$, $D_T$ the link distance $L$ the equations are satisfied. The results of this check are given in Fig.~\ref{fig:4}. From Fig.~\ref{fig:4} we can see that a reasonable range of telescope apertures should provide such a loss. That is, we are indeed emulating the losses of LEO-to-ground transmission (under good conditions).

\section{\label{sec:RnD}Results and Discussion}
We implemented a real-time prototype QKD system integrated with PQC over a channel length of $1.5$~m in a laboratory environment. Channel transmittance was determined by transmitting a $633$~nm beacon laser through the fiber and free-space channel, yielding a measured transmittance of 0.61 (the $4.3$~dB channel loss referred to in the previous section). 
\begin{table*}
    \centering
    \caption{The values for the channel transmission, visibility, QBER, and secure key rate (kbps) at different optical powers of the EPS.}
    \begin{tabular}{|c|cccccc|}
        \hline
        Source power (mW) &  \multicolumn{1}{c|}{Visibility} & \multicolumn{1}{c|}{QBER} & \multicolumn{1}{c|}{Secure key rate (kbps)}
        \\ \hline
        2.5 &  \multicolumn{1}{c|}{$0.960\pm0.009$} & \multicolumn{1}{c|}{0.016$\pm0.005$}& \multicolumn{1}{c|}{1.6}
        \\ \hline
        3.5 & \multicolumn{1}{c|}{$0.873\pm0.019$} & \multicolumn{1}{c|}{$0.063\pm0.009$}& \multicolumn{1}{c|}{0.7}
        \\ \hline
        4.5 & \multicolumn{1}{c|}{$0.742\pm0.013$} & \multicolumn{1}{c|}{$0.136\pm0.008$}& \multicolumn{1}{c|}{Nil}
        \\ \hline
    \end{tabular}    
    \label{table:1}
\end{table*} 
\par We conducted numerous test entanglement measurements for the EPS, altering the source power and measuring visibility~\cite{guo2023high}. We then extracted the QBER, sifted key, and secure key rate for different optical powers of the source. The experimentally measured values are provided in Table \ref{table:1}. The QBER, $Q$, can be calculated directly from visibility using the relation $Q = \frac{1-\text{Visibility}}{2}$, and the secure key rate calculated using Eq.~\ref{eq:key_rate}.
From Table~\ref{table:1}, we observe that the visibility is highest at an optical power of 2.5~mW and decreases as the source power increases. The key rate could only be extracted at lower power levels,  since increasing the source power reduces visibility, resulting in a higher QBER. Since the QBER threshold for a nonzero key rate is 11\%, no secure key can be extracted above this limit. 
\begin{figure}[h]
		\centering		
        \includegraphics[width=01\linewidth]{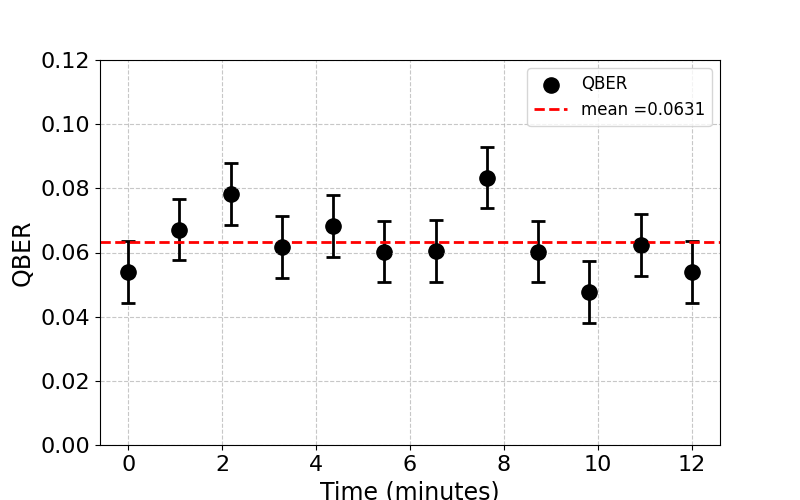}		\caption{The QBER for a duration of 12~minutes.}
 \label{fig:5}
\end{figure}
\par We ran the QKD experiment and measured the live QBER, for a fixed optical power of 3.5~mW to check the stability of the setup. The graph for the estimated live QBER, shown in Fig.~\ref{fig:2}, is extracted for a sifted key length of $1000$ bits (larger block lengths were used in the complete QKD implementation). The figure indicates that the QBER remains stable in different runs and stays within the acceptable threshold limit. This confirms the stability of the background counts (and the overall experiment). From Table~\ref{table:1}, we see that viable secure key rates consistent with what is expected for satellite-to-ground lossy channels are obtained. These QKD keys developed are then used within a hybrid QKD-PQC processing unit to encrypt the data in Alice and decrypt them in Bob, as are syndrome messages within the QKD protocol (as discussed previously).
\par We performed the combined encryption-decryption of QKD-PQC for different sizes of data with our system. In terms of the operation and of the additional processing time associated with the AES-256 encryption at Alice and decryption at Bob, we can confirm that the impact on the processing time (and therefore the real-time QKD key rate) is negligible. This means that additional security can be afforded to the combined system, without any significant cost to the QKD implementation, validating our entire deployed system.
\par However, we should make clear exactly what we mean by `additional' security in this context. If QKD is perfectly implemented and all the assumptions underpinning it are valid in practice, our system does not offer additional security, the presence of AES-256 is inconsequential.  However, if the QKD is compromised, our system does offer additional security in that we at least have PQC, that is, computational security against future quantum computers. We should also be clear that this outcome is not unique - previous work which adopts a different approach, namely solutions based on combining QKD and AES-256 keys before encryption/decryption, offer the same security outcome as ours, provided that at least one of the primitives, QKD or AES, is not compromised in any manner~\cite{garms2024experimental,bruckner2023end}. The difference between these previous combined QKD-PQC systems and that which we offer here will lie in the realm where practical attacks on the primitives are of concern. For example, all QKD protocols are vulnerable to various side-channel attacks, and implementations where the underlying mathematical assumptions from which security is derived are found lacking. Compared to fiber links, these vulnerabilities are likely more prevalent in long-range satellite links. AES-256 is also open to well-known attacks as a consequence of poor implementation, with the related-key attack being the most common \cite{biryukov2009related}. It is under these types of practical attacks on QKD and PQC that the advantages will surface in the different types of combined QKD-PQC approaches. In particular, in schemes such as ours, when all classical reconciliation messages are additionally protected by PQC, new trade-offs in rates versus security will arise. A full theoretical analysis of these differences will form part of our ongoing research in this area. However, we can already note that the AES-256 encryption of syndrome messages increases the QKD rate, albeit only at a post-quantum level of security. That is, if the QKD implementation is trusted, we can use the QKD keys developed for new keys within AES. 
Note that if the AES-256 implementation is trusted, log~$m$ bit of the QKD bits can be saved by eliminating classical authentication. However, full-blown information-theoretic security would only apply to the QKD keys developed under the usual assumptions that the syndrome messages were authenticated classically but heard by the adversary.

It may be worth ending this work with an additional discussion on the different types of QKD systems integrated with PQC currently being discussed in the literature. The difference between these previous combined PQC-QKD systems and that which we offer here will lie in the realm
where practical attacks on the primitives are of concern. Specific attacks will likely lead to different forms of the differences between both approaches and will require a level of detail not presented here. Future work could concentrate on formal analyses of the difference between the different approaches.
We do point out, however, that proposed hybrid key derivation models that combine multiple key primitives from classical key share schemes, post-quantum key share schemes, and QKD schemes require all the keys to be present and of equal size, and the hybrid keys should be secure in a composability context. On the other hand, since our system does not compose the key primitives, the key sizes can be arbitrary, and the security arguments are individually satisfied via a series of encryption, so we do not have to concern ourselves with composable security arguments. 

\section{\label{sec:Conclusion}Conclusion} 
This work reports on an entanglement-based QKD system designed for the satellite-to-Earth channel. The prototype demonstrates the deployment of a combined cryptographic solution that uniquely combines post-quantum security and quantum security, resulting in a system that delivers state-of-the-art communications secure against future quantum computers. We believe that many future implementations of QKD will include post-quantum cryptography solutions in a manner similar to what is outlined here.  Our future work will entail the introduction of advanced tracking systems, enhanced-quantum-enabled network synchronization, finite-sampling effects, space-proofing for the robust environment of low Earth orbit, and a full theoretical analyses of the security advantages of our combined QKD-PQC system under specific attacks.

\section*{\label{sec:Acknowledgments}Acknowledgments}
The authors thank Dr.~Dushy Tissainayagam from Northrop Grumman Australia for constructive feedback and support received during the course of this experiment and Dean Poulos from Quantum Machines for assistance in setting up the signal processing unit used in the experiment. This research has been carried out as a project co-funded by Northrop Grumman Australia, and the Defence Trailblazer Program, a collaborative partnership between the University of Adelaide and the University of New South Wales co-funded by the Australian Government, Department of Education.
\bibliographystyle{IEEEtran}
\bibliography{IEEEabrv,reference}

\end{document}